\documentclass[conference,a4paper]{IEEEtran}
\usepackage[pdftex]{graphicx,color}
\usepackage{curves,epic,latexsym,amsmath,amssymb,array,cite,ifthen,url}
\usepackage{epsfig}

\newcommand{\eqdef}{\stackrel{\scriptscriptstyle\bigtriangleup}{=} }

\newcommand{\Z}{\mathbb{Z}}
\newcommand{\R}{\mathbb{R}}

\newcommand{\calC}{\mathcal{C}}

\newcommand{\lcf}{\operatorname{lcf}}

\newcommand{\wH}{\mathrm{w_H}}

\definecolor{grey}{rgb}{0.5, 0.5, 0.5}

\newcommand{\grey}{\color{grey}}

\newcounter{proglinecounter}
\newenvironment{pseudocode}%
    {\setcounter{proglinecounter}{0}%
     \begin{tabbing}1234\=1234\=123\=123\=123\=123\=123\=123\=123\=123\=123\= \kill}%
    {\end{tabbing}}
\newcommand{\npcl}[1][]
    {\>\refstepcounter{proglinecounter}\arabic{proglinecounter}%
     \ifthenelse{\equal{#1}{}}{}{\label{#1}}\' \>}
\newcommand{\pkw}[1]{\textbf{#1}}    

\newcounter{assertioncntr}
\newcommand{\assertlabel}[1]{\hfill\refstepcounter{assertioncntr}(A.\arabic{assertioncntr})\label{#1}}
\newcommand{\assertref}[1]{A.\ref{#1}}

\newcounter{examplecntr}
{\begin{trivlist}\item[]\refstepcounter{examplecntr}%
 {\bfseries Example~\theexamplecntr%
  \ifthenelse{\equal{#1}{}}{}{ (#1)}.
}}%
{\hfill$\Box$\end{trivlist}}

\newcounter{definitioncntr}
{\begin{trivlist}\item[]\refstepcounter{definitioncntr}%
{\bfseries Definition~\thedefinitioncntr%
 \ifthenelse{\equal{#1}{}}{}{ (#1)}.
}}%
{\hfill$\Box$\end{trivlist}}

\newcounter{theoremcntr}
\newenvironment{theorem}[1][]%
{\begin{trivlist}\item[]\refstepcounter{theoremcntr}%
{\bfseries Theorem~\thetheoremcntr%
  \ifthenelse{\equal{#1}{}}{}{ (#1)}.
}}%
{\hfill$\Box$\end{trivlist}}

\newcounter{lemmacntr}
\newenvironment{lemma}[1][]%
{\begin{trivlist}\item[]\refstepcounter{lemmacntr}%
{\bfseries Lemma~\thelemmacntr%
  \ifthenelse{\equal{#1}{}}{}{ (#1)}.
}}%
{\hfill$\Box$\end{trivlist}}

\newcounter{corollarycntr}
{\begin{trivlist}\item[]\refstepcounter{corollarycntr}%
{\bfseries Corollary~\thecorollarycntr%
  \ifthenelse{\equal{#1}{}}{}{ (#1)}.
}}%
{\hfill$\Box$\end{trivlist}}

\newcounter{propositioncntr}
\newenvironment{proposition}[1][]%
{\begin{trivlist}\item[]\refstepcounter{propositioncntr}%
{\bfseries Proposition~\thepropositioncntr%
  \ifthenelse{\equal{#1}{}}{}{ (#1)}.
}}%
{\hfill$\Box$\end{trivlist}}

\newenvironment{proofof}[1]{\begin{trivlist}\item[]{\bfseries Proof\ifthenelse{\equal{#1}{}}{}{ #1}:}
 }{\hfill$\Box$\end{trivlist}}

\newcommand{\eproofnegspace}{\\[-1.5\baselineskip]\rule{0em}{0ex}}

\begin{document}

\sloppy

\title{Reverse Berlekamp-Massey Decoding}

\author{\IEEEauthorblockN{Jiun-Hung~Yu and~Hans-Andrea~Loeliger}
\IEEEauthorblockA{Department of Information Technology and
Electrical Engineering\\
ETH Zurich, Switzerland\\
Email: \{yu, loeliger\}@isi.ee.ethz.ch}}

\maketitle

\begin{abstract}
We propose a new algorithm for decoding Reed-Solomon codes (up to
half the minimum distance) and for computing inverses in
$F[x]/m(x)$. The proposed algorithm is similar in spirit and
structure to the Berlekamp-Massey algorithm, but it works
naturally for general $m(x)$.
\end{abstract}

\section{Introduction}
\label{sec:Introduction}

In this paper, we propose a new algorithm that solves the following problem.
\begin{trivlist}\item{}{\bfseries Partial-Inverse Problem:}
Let $b(x)$ and $m(x)$ be nonzero polynomials over some finite field $F$,
with $\deg b(x)<\deg m(x)$.
Find a nonzero polynomial $\Lambda(x)\in F[x]$ of the smallest degree such that
\begin{equation} \label{eqn:fundprob}
\deg\Big(b(x)\Lambda(x) \bmod m(x)\Big)< d
\end{equation}
for fixed $d\in\Z$, $1\leq d \leq \deg m(x)$.
\hfill$\Box$
\end{trivlist}

In the special case where $d=1$ and \mbox{$\gcd\!\big(b(x), m(x)\big)=1$},
the problem reduces to computing the inverse of $b(x)$ in $F[x]/m(x)$.

Another special case of the Partial-Inverse Problem
is the standard key equation for decoding Reed-Solomon codes
\cite{Reed,Berlekamp,Sugiyama,Blahut,Roth}
(see Section~\ref{sec:StandardKeyEq}).
In this case, we have $m(x) = x^{n-k}$, where $n$ and $k$ are the blocklength and the dimension
of the code, respectively.
The proposed algorithm then essentially coincides with the Berlekamp-Massey algorithm
\cite{Berlekamp,Massey}
except that it processes the polynomial $b(x)$ (the syndrome) in the reverse order.

The Partial-Inverse Problem with general $m(x)$ arises from an
alternative key equation that will be discussed in
Section~\ref{sec:Decoding}. This alternative key equation and the
corresponding decoding algorithm generalize naturally to
polynomial remainder codes
\cite{Stone,Shiozaki,YuLoeliger,YuLoeliger:prc:ArXiv2012}, which
have not been amenable to Berlekamp-Massey decoding.

The Partial-Inverse Problem can also be solved by a version of the Euclidean algorithm
in the style of \cite{Sugiyama,Roth,Shiozaki,Gao}.
In fact,
it has long been known that the Berlekamp-Massey algorithm and the Euclidean algorithm
are related \cite{Dornstetter,Fitzpatrick,Heydtmann},
and explicit translations were given in \cite{Dornstetter,Heydtmann}.
In this respect,
the algorithm proposed in this paper allows such a translation that is particularly transparent.
However, this topic is not elaborated in the present paper due to lack of space.

The paper is structured as follows.
Section~\ref{sec:RemarksOnProblem} comprises a number of remarks on
the Partial-Inverse Problem, including its application to the standard key equation.
The new algorithm is proposed in Section~\ref{sec:Alg}
and proved in Sections \ref{sec:KeyElements} and~\ref{section:proofAlgo}.
Decoding Reed-Solomon codes via the alternative key equation is described in Section~\ref{sec:Decoding},
and the generalization of this approach to polynomial remainder codes
is outlined in the appendix.

The following notation will be used. The Hamming weight of $e\in
F^n$ will be denoted by $\wH(e)$. The coefficient of $x^\ell$ of a
polynomial $b(x)\in F[x]$ will be denoted $b_\ell$. The leading
coefficient (i.e., the coefficient of $x^{\deg b(x)}$) of a
nonzero polynomial $b(x)$ will be denoted by $\lcf b(x)$, and we
also define $\lcf(0) \eqdef 0$. We will use ``${\bmod}$'' both as
in $r(x) = b(x) \bmod m(x)$ (the remainder of a division) and as
in $b(x) \equiv r(x) \bmod m(x)$ (a congruence modulo $m(x)$). For
$x\in\R$, $\lceil x \rceil$ is the smallest integer not smaller
than $x$.

\section{Remarks}
\label{sec:RemarksOnProblem}

\subsection{General Remarks}

We begin with a number of remarks on the Partial-Inverse Problem as stated
in Section~\ref{sec:Introduction}.
\begin{enumerate}
\item The stated assumptions imply $\deg m(x)\geq 1$.

\item For $d=\deg m(x)$, the problem is solved by $\Lambda(x)=1$.
Smaller values of $d$ will normally require a polynomial
$\Lambda(x)$ of higher degree. \item \label{enumi:RemarkInverse}
In the special case where $d=1$, we have the following solutions.
If $\gcd\!\big(b(x), m(x)\big)=1$, then $b(x)$ has an inverse in
$F[x]/m(x)$ and $\Lambda(x)$ is that inverse (up to a scale
factor); otherwise, the solution is $\Lambda(x) = m(x) /
\gcd\!\big(b(x), m(x)\big)$, which yields $b(x)\Lambda(x) \bmod
m(x) = 0$.

\item The previous remark implies that the problem has a solution
for any $d\geq 1$.

\item We will see that the solution $\Lambda(x)$ of the problem is
unique up to a scale factor (Proposition~\ref{propo:Uniqueness} in
Section~\ref{sec:KeyElements}) and satisfies
\begin{equation} \label{eqn:MaxDegree}
\deg \Lambda(x) \leq \deg m(x) - d
\end{equation}
(by~(\ref{eqn:DegBoundLambdaRet}) in Section~\ref{section:proofAlgo}).

\item
In consequence of (\ref{eqn:MaxDegree}), coefficients $b_\ell$ of $b(x)$ with
\begin{equation} \label{eqn:IrrelevantCoeff_b}
\ell < 2d - \deg m(x)
\end{equation}
and coefficients $m_\ell$ of $m(x)$ with
\begin{equation} \label{eqn:IrrelevantCoeff_m}
\ell < 2d - \deg m(x) + 1
\end{equation}
are irrelevant for the solution $\Lambda(x)$:
these coefficients do not affect (\ref{eqn:fundprob}) since
\begin{equation}
b(x) \Lambda(x) \bmod m(x) = b(x) \Lambda(x) - q(x) m(x)
\end{equation}
with $\deg q(x) < \deg \Lambda(x) \leq \deg m(x) - d$.

Such irrelevant coefficients may be set to zero without affecting the solution $\Lambda(x)$.
\end{enumerate}

\subsection{Application to the Standard Key Equation}
\label{sec:StandardKeyEq}

The standard key equation for decoding Reed-Solomon codes
\cite{Berlekamp,Sugiyama,Fitzpatrick,Blahut,Roth} is
\begin{equation} \label{eqn:classicalkey}
S(x) \Lambda(x) \equiv \Gamma(x) \bmod x^{n-k},
\end{equation}
where $n$ and $k$ are the blocklength and the dimension of the code, respectively,
and where $S(x)$ is a (given) syndrome polynomial with $\deg S(x)<n-k$.
The desired solution is a pair $\Gamma(x)$ and $\Lambda(x)\neq 0$
such that $\deg \Gamma(x) < \deg \Lambda(x) \leq (n-k)/2$.

The problem of finding such a pair $\Gamma(x)$ and $\Lambda(x)$
translates into a Partial-Inverse Problem
with \mbox{$b(x)=S(x)$}, \mbox{$m(x)=x^{n-k}$}, and \mbox{$d=\lceil (n-k)/2 \rceil$}.
Because of (\ref{eqn:MaxDegree}), the resulting $\Lambda(x)$
satisfies $\deg \Lambda(x) \leq (n-k)/2$,
and we have $\Gamma(x) = S(x)\Lambda(x) \bmod x^{n-k}$.
If the number of errors does not exceed $(n-k)/2$, the condition $\deg \Gamma(x) < \deg \Lambda(x)$
will then be satisfied automatically.

\section{The Algorithm}
\label{sec:Alg}

The Partial-Inverse Problem as stated in Section~\ref{sec:Introduction}
can be solved by the following algorithm.

\begin{trivlist}
\item{}{\bf Proposed Algorithm:}\\
\textbf{Input:} $b(x)$, $m(x)$, and $d$ as in the problem statement.\\
\textbf{Output:}
$\Lambda(x)$ as in the problem statement.
\begin{pseudocode}
\npcl[line:2noErrorBegin] \pkw{if} $\deg b(x)< d$ \pkw{begin} \\
\npcl[line:2noError] \> \pkw{return} $\Lambda(x):=1$\\
\npcl[line:2noErrorEnd] \pkw{end} \\
\npcl[line:2initLambda1] $\Lambda^{(1)}(x) := 0,\ d_1 := \deg m(x),\ \kappa_1:=\lcf m(x)$ \\
\npcl[line:2initLambda2] $\Lambda^{(2)}(x) := 1,\ d_2 := \deg b(x),\ \kappa_2:=\lcf b(x)$ \\
\npcl[line:2loopbegin]   \pkw{loop begin} \\
\npcl[line:2updateLambda1] \>
$\Lambda^{(1)}(x):= \kappa_2 \Lambda^{(1)}(x)- \kappa_1 x^{d_1-d_2} \Lambda^{(2)}(x)$ \\
\> \> \> {\grey\rule[0.5ex]{50mm}{1pt}}\\
\npcl[line:2updated1] \>$d_1:=\deg \left(b(x)\Lambda^{(1)}(x)\bmod m(x)\right)$\\
\npcl[line:2ifbegin] \> \pkw{if} $d_1< d$ \pkw{begin} \\
\npcl[line:2return] \>\>\pkw{return} $\Lambda(x) := \Lambda^{(1)}(x)$\\
\npcl[line:2ifend]\> \pkw{end} \\
\npcl[line:2updatekap1] \> $\kappa_1:=\lcf\big(b(x)\Lambda^{(1)}(x)\bmod m(x)\big)$\\
\> \> \> {\grey\rule[0.5ex]{50mm}{1pt}}\\
\npcl[line:2ifd1d2begin] \> \pkw{if} $d_1<  d_2$ \pkw{begin} \\
\npcl[line:2swapbegin] \>\>$(\Lambda^{(1)}(x),\Lambda^{(2)}(x)):=(\Lambda^{(2)}(x),\Lambda^{(1)}(x))$\\
\npcl \>\>$(d_1, d_2):=(d_2,d_1)$\\
\npcl[line:2swapend] \>\>$(\kappa_1, \kappa_2):=(\kappa_2, \kappa_1)$\\
\npcl[line:2ifd1d2end] \> \pkw{end} \\
\npcl[line:2loopend]   \pkw{end}\\
\end{pseudocode}
\vspace{-6ex} \hfill $\Box$
\end{trivlist}

Note that lines \ref{line:2swapbegin}--\ref{line:2swapend} simply
swap $\Lambda^{(1)}(x)$ with $\Lambda^{(2)}(x)$, $d_1$ with $d_2$,
and $\kappa_1$ with $\kappa_2$.
The only actual computations are in lines \ref{line:2updateLambda1} and~\ref{line:2updated1}.

The correctness of this algorithm will be proved in
Section~\ref{section:proofAlgo}.
In particular, we will see
that the value of $d_1$ is reduced in every execution of
line~\ref{line:2updated1}.

Note that lines~\ref{line:2updated1} and~\ref{line:2updatekap1}
do not require the computation of the entire polynomial $b(x) \Lambda^{(1)}(x)\bmod m(x)$.
Indeed, lines~\ref{line:2updated1}--\ref{line:2updatekap1} can be replaced by the following loop:
\begin{pseudocode}
\textbf{Equivalent Alternative to Lines~\ref{line:2updated1}--\ref{line:2updatekap1}:}\\[1ex]
\setcounter{proglinecounter}{30}
\npcl \> \pkw{repeat}\\
\npcl \> \> $d_1 := d_1 - 1$ \\
\npcl \> \> \pkw{if $d_1< d$ begin} \\
\npcl \> \> \> \pkw{return} $\Lambda(x) := \Lambda^{(1)}(x)$\\
\npcl \> \> \pkw{end} \\
\npcl[kappa1alt] \> \> $\kappa_1 := \text{coefficient of $x^{d_1}$ in}$ \\
      \> \> \> \> \> \> $b(x) \Lambda^{(1)}(x) \bmod m(x)$ \\
\npcl \> \pkw{until} $\kappa_1 \neq 0$
\end{pseudocode}

In the special case where $m(x)=x^\nu$, line~\ref{kappa1alt} amounts to
\begin{pseudocode}
\setcounter{proglinecounter}{40} \npcl[line:2altupdateLam1xn] \>
$\kappa_1 :=
b_{d_1} \Lambda_0^{(1)} + b_{d_1-1} \Lambda_1^{(1)} + \ldots + b_{d_1-\tau} \Lambda_\tau^{(1)}$
\end{pseudocode}
with $\tau \eqdef \deg \Lambda^{(1)}(x)$ and where $b_\ell \eqdef 0$ for $\ell<0$.
In the other special case where $m(x)=x^n-1$ as in (\ref{eqn:xn1}) below,
line~\ref{kappa1alt} becomes
\begin{pseudocode}
\setcounter{proglinecounter}{50} \npcl[line:2altupdateLam1xn1] \>
$\kappa_1 :=
b_{d_1} \Lambda_0^{(1)} + b_{[d_1-1]} \Lambda_1^{(1)} + \ldots + b_{[d_1-\tau]} \Lambda_\tau^{(1)}$
\end{pseudocode}
with $b_{[\ell]} \eqdef b_{\ell \bmod n}$.
In both cases, the proposed algorithm looks very much like, and is as efficient as,
the Berlekamp-Massey algorithm \cite{Massey}.

\section{Decoding Reed-Solomon Codes via an Alternative Key Equation}
\label{sec:Decoding}

Decoding Reed-Solomon codes (up to half the minimum distance)
can be reduced rather directly to the Partial-Inverse Problem of Section~\ref{sec:Introduction}
as follows.

Let $F$ be a finite field,
let $\beta_0,\ldots,\beta_{n-1}$ be $n$ different elements of $F$,
let $m(x) \eqdef \prod_{\ell=0}^{n-1} (x-\beta_\ell)$,
let $F[x]/m(x)$ be the ring of polynomials modulo $m(x)$,
and let $\psi$ be the evaluation mapping
\begin{equation} \label{eqn:DefMappingPsi}
\psi: F[x]/m(x) \rightarrow F^n: a(x) \mapsto \big( a(\beta_0),\ldots, a(\beta_{n-1}) \big),
\end{equation}
which is a ring isomorphism. A Reed-Solomon code with blocklength
$n$ and dimension $k$ may be defined as
\begin{equation} \label{eqn:DefGenRS}
\{ c=(c_0,\ldots,c_n) \in F^n: \deg \psi^{-1}(c) < k \},
\end{equation}
usually with the additional condition that
\begin{equation} \label{eqn:EvalPointsPowersAlpha}
\beta_\ell = \alpha^{\ell} \text{~~~for $\ell=0,\ldots,n-1$},
\end{equation}
where $\alpha\in F$ is a primitive $n$-th root of unity.
The condition (\ref{eqn:EvalPointsPowersAlpha})
implies
\begin{equation} \label{eqn:xn1}
m(x) = x^n-1
\end{equation}
and turns $\psi$ into a discrete Fourier transform
\cite{Blahut}.
However, (\ref{eqn:EvalPointsPowersAlpha}) and (\ref{eqn:xn1}) will not be required below.

Let $y=(y_0,\ldots,y_{n-1}) \in F^n$ be the received word,
which we wish to decompose into
\begin{equation}
y = c+e
\end{equation}
where $c\in \calC$ is a codeword
and where the Hamming weight of $e=(e_0,\ldots,e_{n-1}) \in F^n$ is as small as possible.

Let $C(x) \eqdef \psi^{-1}(c)$, and analogously
$E(x) \eqdef \psi^{-1}(e)$ and $Y(x) \eqdef \psi^{-1}(y)$.
Clearly, we have
$\deg C(x)<k$ and $\deg E(x) < \deg m(x) = n$.

For any $e\in F^n$, we define the error locator polynomial
\begin{equation} \label{eqn:el}
\Lambda_e(x) \eqdef \prod_{\scriptsize\begin{array}{cc} \ell\in
\{0,\ldots, n-1\} \\  e_\ell\neq 0 \end{array}} (x-\beta_\ell).
\end{equation}
Clearly, $\deg \Lambda_e(x) = \wH(e)$ and
\begin{equation} \label{eqn:KeyEq}
E(x) \Lambda_e(x) \bmod m(x) = 0.
\end{equation}

\begin{theorem}[Alternative Key Equation] \label{theorem:OurKey}
If $\wH(e) \leq \frac{n-k}{2}$, then
the error locator polynomial $\Lambda_e(x)$ satisfies
\begin{equation} \label{eqn:OurKeyLambdae}
\deg \big( Y(x) \Lambda_e(x) \bmod m(x) \big) < n-\frac{n-k}{2}
\end{equation}
Conversely, for any $y$ and $e\in F^n$ and $t\in\R$ with
\begin{equation} \label{eqn:OurKeyt}
\wH(e) \leq t \leq \frac{n-k}{2},
\end{equation}
if some nonzero $\Lambda(x)\in F[x]$ with $\deg \Lambda(x) \leq t$
satisfies
\begin{equation} \label{eqn:OurKey}
\deg \Big( Y(x) \Lambda(x) \bmod m(x) \Big) < n-t,
\end{equation}
then $\Lambda(x)$ is a multiple of $\Lambda_e(x)$.
\end{theorem}
The proof is not difficult, but omitted due to lack of space.
We thus arrive at the following decoding procedure:
\begin{enumerate}
\item
Compute $Y(x)=\psi^{-1}(y)$.
\item
Run the algorithm of Section~\ref{sec:Alg}
with $b(x) = Y(x)$ and
$d = \lceil\frac{n+k}{2}\rceil$.
If $\wH(e) \leq \frac{n-k}{2}$, then the polynomial $\Lambda(x)$
returned by the algorithm equals $\Lambda_e(x)$ up to a scale factor.
\item
Complete decoding by any standard method
\cite{Blahut} or by means of Proposition~\ref{theorem:RecoverCxByMultDiv} below.
\end{enumerate}

Note that in Step~2, because of (\ref{eqn:IrrelevantCoeff_b}),
coefficients $Y_\ell$ of $Y(x)$ with
\begin{equation}
\ell <  \ell_\text{min}
  \eqdef \left\{ \begin{array}{ll}
         k, & \text{if $n-k$ is even} \\
         k+1, & \text{if $n-k$ is odd}
        \end{array}\right.
\end{equation}
are irrelevant for finding $\Lambda(x)$ and can be set to zero.
The remaining coefficients $Y_\ell$ are syndromes since $C_\ell=0$ and $Y_\ell = E_\ell$
for $\ell\geq \ell_\text{min}$.

As mentioned, decoding can be completed by the following proposition:
\begin{proposition} \label{theorem:RecoverCxByMultDiv}
If $\Lambda(x)$ is a nonzero multiple of $\Lambda_e(x)$ with $\deg \Lambda(x) \leq n-k$,
then
\begin{equation} \label{eqn:RecoverCxByMultDiv}
C(x) = \frac{Y(x) \Lambda(x) \bmod m(x)}{\Lambda(x)}
\end{equation}
\eproofnegspace
\end{proposition}
\begin{proofof}{}
If $\Lambda(x)$ has the stated properties, then
\begin{IEEEeqnarray}{rCl}
\IEEEeqnarraymulticol{3}{l}{
Y(x) \Lambda(x) \bmod m(x)
}\nonumber\\\quad
 & = & C(x) \Lambda(x) \bmod m(x) + E(x) \Lambda(x) \bmod m(x)
       \IEEEeqnarraynumspace \label{eqn:ProofRecoverCxByMultDiv}\\
 & = & C(x) \Lambda(x),  \label{eqn:YxLambdaxModmCxLambdax}
\end{IEEEeqnarray}
where the second term in (\ref{eqn:ProofRecoverCxByMultDiv})
vanishes because of (\ref{eqn:KeyEq}).
\end{proofof}

Note that computing the numerator of (\ref{eqn:RecoverCxByMultDiv})
may be viewed as continuing the algorithm of Section~\ref{sec:Alg}
(line~\ref{kappa1alt}) with frozen $\Lambda^{(1)}(x) = \Lambda(x)$.

\section{Key Elements of the Proof}
\label{sec:KeyElements}

We now turn to the proof of the algorithm proposed in
Section~\ref{sec:Alg}. In this section, we discuss some key
elements of the proof; the actual proof will then be given in
Section~\ref{section:proofAlgo}.

The pivotal part of the algorithm is
line~\ref{line:2updateLambda1}, which is explained by the
following simple lemma. (The corresponding statement for the
Berlekamp-Massey algorithm is the \emph{two-wrongs-make-a-right}
lemma, so called by J.~L.~Massey.)

\begin{lemma} \label{lemma:Algcore}
Let $m(x)$ be a polynomial over $F$ with $\deg m(x) \geq 1$. For
further polynomials $b(x), \Lambda^{(1)}(x), \Lambda^{(2)}(x) \in
F[x]$, let
\begin{IEEEeqnarray}{rCl}
r^{(1)}(x) & \eqdef & b(x) \Lambda^{(1)}(x) \bmod m(x), \label{eqn:CoreLemmaR1} \IEEEeqnarraynumspace\\
r^{(2)}(x) & \eqdef & b(x) \Lambda^{(2)}(x) \bmod m(x), \label{eqn:CoreLemmaR2}
\end{IEEEeqnarray}
$d_1 \eqdef \deg r^{(1)}(x)$, $\kappa_1 \eqdef \lcf r^{(1)}(x)$,
$d_2 \eqdef \deg r^{(2)}(x)$, $\kappa_2 \eqdef \lcf r^{(2)}(x)$,
and assume $d_1 \geq d_2 \geq 0$.
Then
\begin{equation} \label{eqn:2WrongsMakeRight}
\Lambda(x) \eqdef \kappa_2 \Lambda^{(1)}(x)- \kappa_1 x^{d_1-d_2} \Lambda^{(2)}(x)
\end{equation}
satisfies
\begin{equation}
\deg \Big( b(x) \Lambda(x) \bmod m(x) \Big) < d_1.
\end{equation}
\eproofnegspace
\end{lemma}

\begin{proofof}{}
From (\ref{eqn:2WrongsMakeRight}), we obtain
\begin{IEEEeqnarray}{rCl}
r(x)  & \eqdef &  b(x) \Lambda(x) \bmod m(x) \IEEEeqnarraynumspace\\
  & = & \kappa_2 r^{(1)}(x) - \kappa_1 x^{d_1-d_2} r^{(2)}(x) \label{eqn:2WrongsMakeRightRemainder}
\end{IEEEeqnarray}
by the natural ring homomorphism $F[x] \rightarrow F[x]/m(x)$.
It is then obvious from (\ref{eqn:2WrongsMakeRightRemainder}) that
$\deg r(x) < \deg r^{(1)}(x)=d_1$.
\end{proofof}

A similar argument proves
\begin{proposition}[Uniqueness of Solution]\label{propo:Uniqueness}
The solution $\Lambda(x)$ of the Partial-Inverse Problem of
Section~\ref{sec:Introduction} is unique up to a scale factor.
\end{proposition}

\begin{proofof}{}
Let $\Lambda^{(1)}(x)$ and $\Lambda^{(2)}(x)$ be two solutions of
the problem, which implies $\deg \Lambda^{(1)}(x)=\deg
\Lambda^{(2)}(x) \geq 0$. Define $r^{(1)}(x)$ and $r^{(2)}(x)$ as
in (\ref{eqn:CoreLemmaR1}) and (\ref{eqn:CoreLemmaR2}) and
consider
\begin{equation} \label{eqn:ProofUniquenessLambda3}
\Lambda(x)\eqdef \Big(\lcf \Lambda^{(2)}(x) \Big) \Lambda^{(1)}(x) - \Big(\lcf \Lambda^{(1)}(x) \Big) \Lambda^{(2)}(x).
\end{equation}
Then
\begin{IEEEeqnarray}{rCl}
r(x)  & \eqdef &  b(x) \Lambda(x) \bmod m(x) \IEEEeqnarraynumspace\\
  & = & \Big(\lcf \Lambda^{(2)}(x) \Big) r^{(1)}(x) - \Big(\lcf \Lambda^{(1)}(x) \Big) r^{(2)}(x),
         \IEEEeqnarraynumspace
\end{IEEEeqnarray}
which implies that $\Lambda(x)$ also satisfies (\ref{eqn:fundprob}).
But (\ref{eqn:ProofUniquenessLambda3}) implies
$\deg \Lambda(x) < \deg\Lambda^{(1)}(x)$,
which is a contradiction unless $\Lambda(x)=0$.
Thus $\Lambda(x)=0$, which means that
$\Lambda^{(1)}(x)$ and $\Lambda^{(2)}(x)$
are equal up to a scale factor.
\end{proofof}

\begin{trivlist}\item{}{\bfseries Definition (Minimal Partial Inverse):}
For fixed nonzero $b(x)$ and \mbox{$m(x)\in F[x]$} with
\mbox{$\deg b(x) < \deg m(x)$}, a nonzero polynomial
$\Lambda(x)\in F[x]$ is a \emph{minimal partial inverse} of $b(x)$
if\vspace{-1ex}
\begin{equation} \label{eqn:DefMinRemPoly}
\deg \Big( b(x) \Lambda^{(1)}(x) \bmod m(x) \Big) \leq \deg \Big( b(x) \Lambda(x) \bmod m(x) \Big)
\end{equation}
(with $\Lambda^{(1)}(x)\neq 0$)
implies $\deg \Lambda^{(1)}(x) \geq \deg \Lambda(x)$.
\hfill$\Box$
\end{trivlist}

The following lemma is the counterpart to Theorem~1
of~\cite{Massey}.
\begin{lemma}[Degree Change Lemma] \label{lemma:DegreeChangeLemma}
For fixed nonzero $b(x)$ and $m(x)\in F[x]$ with $\deg b(x) < \deg
m(x)$, let $\Lambda(x)$ be a minimal partial inverse of $b(x)$ and
let
\begin{equation} \label{eqn:LemmaMinimalRemainder}
r(x) \eqdef  b(x) \Lambda(x) \bmod m(x).
\end{equation}
If
\begin{equation} \label{eqn:DegreeChangeLemmaIf}
\deg \Lambda(x) \leq \deg m(x) - \deg r(x),
\end{equation}
then any nonzero polynomial $\Lambda^{(1)}(x)\in F[x]$ such that
\begin{equation}  \label{theorem:minsoltmp}
\deg \big( b(x)\Lambda^{(1)}(x) \bmod m(x) \big) < \deg r(x)
\end{equation}
satisfies
\begin{equation} \label{theorem:minsoldeg}
\deg \Lambda^{(1)}(x)\geq  \deg m(x)-\deg r(x).
\end{equation}
\eproofnegspace
\end{lemma}
The proof is given below.

\begin{trivlist}\item{}{\bfseries Corollary:}
Assume everything as in Lemma~\ref{lemma:DegreeChangeLemma}
including (\ref{eqn:DegreeChangeLemmaIf}) and
(\ref{theorem:minsoltmp}). If (\ref{theorem:minsoldeg}) is
satisfied with equality, then $\Lambda^{(1)}(x)$ is also a minimal
partial inverse of $b(x)$.
\end{trivlist}

\begin{proofof}{of Lemma~\ref{lemma:DegreeChangeLemma}}
Assume that $\Lambda^{(1)}(x)$ is a nonzero polynomial
that satisfies (\ref{theorem:minsoltmp}),
i.e., the degree of
\begin{equation} \label{lemma:MinimalRemainder2}
r^{(1)}(x) \eqdef  b(x) \Lambda^{(1)}(x) \bmod m(x)
\end{equation}
satisfies
\begin{equation} \label{eqn:ProofDCLdegrr}
\deg r^{(1)}(x) < \deg r(x).
\end{equation}
Multiplying (\ref{eqn:LemmaMinimalRemainder}) by $\Lambda^{(1)}(x)$
and (\ref{lemma:MinimalRemainder2}) by $\Lambda(x)$
yields
\begin{equation} \label{eqn:ProofDCLprod}
\Lambda^{(1)}(x) r(x) \equiv \Lambda(x) r^{(1)}(x) \mod m(x).
\end{equation}
If we assume both (\ref{eqn:DegreeChangeLemmaIf})
and (contrary to (\ref{theorem:minsoldeg}))
\begin{equation} \label{proof:assmindeg}
\deg \Lambda^{(1)}(x)< \deg m(x)-\deg r(x),
\end{equation}
then (\ref{eqn:ProofDCLprod}) reduces to
\begin{equation}\label{proof:minsoleqn}
\Lambda^{(1)}(x)r(x)= \Lambda(x) r^{(1)}(x).
\end{equation}
But then (\ref{eqn:ProofDCLdegrr}) implies $\deg \Lambda^{(1)}(x)<
\deg \Lambda(x)$, which is impossible because $\Lambda(x)$ is a
minimal partial inverse. Thus (\ref{eqn:DegreeChangeLemmaIf}) and
(\ref{proof:assmindeg}) cannot hold simultaneously.
\end{proofof}

\section{Proof of the Proposed Algorithm}
\label{section:proofAlgo}

We now prove the correctness of the algorithm proposed in
Section~\ref{sec:Alg}. To this end, we restate
the algorithm with added assertions as follows.

\begin{pseudocode}
\textbf{Proposed Algorithm Restated:}\\[1ex]
\npcl[line:pnoErrorBegin] \pkw{if} $\deg b(x) < d$ \pkw{begin} \\
\npcl[line:pnoError] \> \pkw{return} $\Lambda(x):=1$\\
\npcl[line:pnoErrorEnd] \pkw{end} \\
\npcl[line:pinitLambda1] $\Lambda^{(1)}(x) := 0,\ d_1 := \deg m(x),\ \kappa_1:=\lcf m(x)$ \\
\npcl[line:pinitLambda2] $\Lambda^{(2)}(x) := 1,\ d_2 := \deg b(x),\ \kappa_2:=\lcf b(x)$ \\
\npcl[line:ploopbegin]   \pkw{loop begin} \\
\> \> \> \framebox[0.75\linewidth]{\begin{minipage}{0.72\linewidth}%
           \pkw{Assertions:}\\
           $d_1 > d_2 \geq d$  \assertlabel{assert:d1Ld2}\\
           $\deg \Lambda^{(2)}(x) = \deg m(x) - d_1$ \assertlabel{assert:loopinvariant.1} \\
           \phantom{$\deg \Lambda^{(2)}(x)$} $ > \deg \Lambda^{(1)}(x)$ \assertlabel{assert:Lam1SLam2}\\
           $\Lambda^{(2)}(x)$ is a minimal partial inverse \assertlabel{assert:Lambda2MinRemPoly}
          \end{minipage}}\\[0.5ex]
\npcl[line:prepeatbegin] \> \pkw{repeat} \\
\npcl[line:pupdateLambda1] \>\>
$\Lambda^{(1)}(x):= \kappa_2 \Lambda^{(1)}(x)- \kappa_1 x^{d_1-d_2} \Lambda^{(2)}(x)$ \\
\> \> \> \>\framebox[0.69\linewidth]{\begin{minipage}{0.66\linewidth}%
           \pkw{Assertions:}\\
            $\deg (b(x)\Lambda^{(1)}(x)\bmod m(x))<d_1$ \assertlabel{assert:Pros1} \\
            $\deg \Lambda^{(1)}(x) = \deg m(x) - d_2$ \assertlabel{assert:loopinvariant.2} \\
            \phantom{$\deg \Lambda^{(1)}(x)$} $ > \deg \Lambda^{(2)}(x)$ \assertlabel{assert:Lam1LLam2}
          \end{minipage}}\\[0.5ex]
\npcl[line:pupdated1] \>\>$d_1:=\deg \left(b(x)\Lambda^{(1)}(x)\bmod m(x)\right)$\\
\npcl[line:pifbegin] \> \> \pkw{if} $d_1 < d$ \pkw{begin} \\
\> \> \> \> \> \framebox[0.63\linewidth]{\begin{minipage}{0.60\linewidth}%
           \pkw{Assertion:}\\
            $\Lambda^{(1)}(x)$ is a min.\ partial inverse \assertlabel{assert:Output}
          \end{minipage}}\\[0.5ex]
\npcl[line:preturn] \> \>\>\pkw{return} $\Lambda(x):=\Lambda^{(1)}(x)$\\
\npcl[line:pifend] \> \> \pkw{end} \\
\npcl[line:pupdatekap1] \>\> $\kappa_1:=\lcf\big(b(x)\Lambda^{(1)}(x)\bmod m(x)\big)$\\
\npcl[line:prepeatend] \> \pkw{until} $d_1 < d_2$ \\
\> \> \> \framebox[0.75\linewidth]{\begin{minipage}{0.72\linewidth}%
           \pkw{Assertion:}\\
            $\Lambda^{(1)}(x)$ is a minimal partial inverse \assertlabel{assert:Lambda1MinRemPoly}
          \end{minipage}}\\[0.5ex]
\npcl[line:pswapbegin] \>$(\Lambda^{(1)}(x),\Lambda^{(2)}(x)):=(\Lambda^{(2)}(x),\Lambda^{(1)}(x))$\\
\npcl \>$(d_1, d_2):=(d_2,d_1)$\\
\npcl[line:pswapend] \>$(\kappa_1, \kappa_2):=(\kappa_2, \kappa_1)$\\
\npcl[line:ploopend]   \pkw{end}\\
\end{pseudocode}

Note the added inner \pkw{repeat} loop (lines \ref{line:prepeatbegin}--\ref{line:prepeatend}),
which does not change the algorithm but helps to state its proof.

Throughout the algorithm (except at the very beginning, before the
first execution of lines \ref{line:pupdated1} and
\ref{line:pupdatekap1}), $d_1$, $d_2$, $\kappa_1$, and $\kappa_2$
are defined as in Lemma~\ref{lemma:Algcore}, i.e., \mbox{$d_1 =
\deg r^{(1)}(x)$}, \mbox{$\kappa_1 = \lcf r^{(1)}(x)$}, \mbox{$d_2
= \deg r^{(2)}(x)$}, and \mbox{$\kappa_2 = \lcf r^{(2)}(x)$} for
$r^{(1)}(x)$ and $r^{(2)}(x)$ as in (\ref{eqn:CoreLemmaR1}) and
(\ref{eqn:CoreLemmaR2}).

Assertions
(\assertref{assert:d1Ld2})--(\assertref{assert:Lambda2MinRemPoly})
are easily verified, both from the initialization and from
(\assertref{assert:loopinvariant.2}),
(\assertref{assert:Lam1LLam2}), and
(\assertref{assert:Lambda1MinRemPoly}).

As for (\assertref{assert:Pros1}), after the very first execution of line~\ref{line:pupdateLambda1},
we still have $d_1=\deg m(x)$ (from line~\ref{line:pinitLambda1}),
which makes (\assertref{assert:Pros1}) obvious.
For all later executions of line~\ref{line:pupdateLambda1},
(\assertref{assert:Pros1}) follows from Lemma~\ref{lemma:Algcore}.

As for (\assertref{assert:loopinvariant.2}) and
(\assertref{assert:Lam1LLam2}), we note that
line~\ref{line:pupdateLambda1} changes the degree of
$\Lambda^{(1)}(x)$ as follows:
\begin{itemize}
\item
Upon entering the \pkw{repeat} loop,
line~\ref{line:pupdateLambda1} increases the degree of $\Lambda^{(1)}$ to
\begin{IEEEeqnarray}{rCl}
\deg \Lambda^{(2)}(x) + d_1 - d_2  & = & \deg m(x) - d_2  \label{eqn:AlgProofDegChange} \IEEEeqnarraynumspace\\
  & > & \deg \Lambda^{(2)}(x), \label{eqn:AlgProofDegChange2}
\end{IEEEeqnarray}
which follows from (\assertref{assert:d1Ld2})--(\assertref{assert:Lam1SLam2}).

\item
Subsequent executions of line~\ref{line:pupdateLambda1}
without leaving the \pkw{repeat} loop
(i.e., without executing lines~\ref{line:pswapbegin}--\ref{line:pswapend})
do not change the degree of $\Lambda^{(1)}(x)$.
(This follows from the fact that
$d_1$ is smaller than in the first execution
while $\Lambda^{(2)}(x)$, $d_2$, and $\kappa_2\neq 0$ remain unchanged.)
\end{itemize}

Assertion (\assertref{assert:Lambda1MinRemPoly}) follows from the
Corollary to Lemma~\ref{lemma:DegreeChangeLemma} (with $\Lambda(x)
= \Lambda^{(2)}(x)$ and $\deg r(x) = d_2$), which applies because
$d_1<d_2$ and (\assertref{assert:loopinvariant.2}).
Because of (\assertref{assert:d1Ld2}),
the same argument applies also to (\assertref{assert:Output}).

Finally, (\assertref{assert:d1Ld2}) and
(\assertref{assert:loopinvariant.2}) imply that the polynomial
$\Lambda(x)$ returned by the algorithm satisfies
\begin{equation} \label{eqn:DegBoundLambdaRet}
\deg \Lambda(x) \leq \deg m(x) - d.
\end{equation}

\section{Conclusion}

We have proposed a new algorithm for decoding Reed-Solomon codes and
polynomial remainder codes, and for computing inverses in $F[x]/m(x)$.
In the special case where $m(x)=x^\nu$ or $m(x)=x^n-1$, the proposed
algorithm almost coincides with the Berlekamp-Massey algorithm,
except that it processes the syndrome in reverse order.

\section*{Appendix: Extension to Polynomial\\ Remainder Codes}

Polynomial remainder codes
\cite{Stone,YuLoeliger,YuLoeliger:prc:ArXiv2012,Shiozaki} are a
class of codes that include Reed-Solomon codes as a special case.
We briefly outline how decoding via the alternative key equation of Section~\ref{sec:Decoding}
generalizes to polynomial remainder codes, which can thus be decoded by the algorithm
of Section~\ref{sec:Alg}.

Let $m_0(x),\ldots,m_{n-1}(x)\in F[x]$ be relatively prime
and let $m(x) \eqdef \prod_{\ell=0}^{n-1} m_\ell(x)$.
Let $R_{m}\eqdef F[x]/m(x)$ denote the ring of
polynomials modulo $m(x)$ and let $R_{m_\ell}\eqdef F[x]/m_\ell(x)$.
The mapping (\ref{eqn:DefMappingPsi})
is generalized to the ring isomorphism
\begin{IEEEeqnarray}{rCl}
   \psi &:& R_{m} \rightarrow R_{m_0} \times \ldots \times R_{m_{n-1}}: \nonumber\\
    &&  a(x) \mapsto \psi(a) \eqdef \big( \psi_0(a),\ldots,\psi_{n-1}(a) \big)
        \label{eqn:defPhi}
\end{IEEEeqnarray}
with $\psi_\ell(a) \eqdef a(x) \bmod m_{\ell}(x)$. Following
\cite{YuLoeliger:prc:ArXiv2012}, a polynomial remainder code may
be defined as
\begin{equation} \label{eqn:DefPRC}
  \{ c=(c_0,\ldots,c_{n-1}) \in  R_{m_0} \times \ldots \times R_{m_{n-1}}:
    \deg \psi^{-1}(c) < K \}
\end{equation}
where
\begin{equation}
K \eqdef \sum_{\ell=0}^{k-1}\deg m_\ell(x)
\end{equation}
for some fixed $k$, $0<k<n$.
We also define
\begin{equation}
N \eqdef \deg m(x) = \sum_{\ell=0}^{n-1}\deg m_\ell(x).
\end{equation}

As in Section~\ref{sec:Decoding}, let $y=c+e$ be the received word
with $c\in \calC$, and let $C(x) \eqdef \psi^{-1}(c)$,
$E(x) \eqdef \psi^{-1}(e)$,
and $Y(x) \eqdef \psi^{-1}(y)$.
Clearly, $\deg C(x)<K$ and $\deg E(x) < N$.

For such codes, the error locator polynomial
\begin{equation}
\Lambda_e(x) \eqdef \prod_{\scriptsize\begin{array}{cc} \ell\in
\{0,\ldots, n-1\} \\  e_\ell\neq 0 \end{array}} m_\ell(x)
\end{equation}
\newpage\noindent
and the error factor polynomial \cite{YuLoeliger:prc:ArXiv2012}
\begin{equation} \label{eqn:ef}
\Lambda_f(x)\eqdef m(x)/\gcd\!\big(E(x),m(x)\big)
\end{equation}
do not, in general, coincide. However, if
all moduli $m_\ell(x)$ are irreducible, then $\Lambda_f(x)=\Lambda_e(x)$.

We then have the following generalization of
Theorem~\ref{theorem:OurKey}:

\begin{theorem} \label{theorem:OurPRCKey}
For given $y$ and $e$ with
$\deg \Lambda_f(x) \leq t \leq \frac{N-K}{2}$,
assume that some nonzero
polynomial $\Lambda(x)$ with $\deg \Lambda(x) \leq t$ satisfies
\begin{equation} \label{eqn:OurPRCKey}
\deg \big( Y(x) \Lambda(x) \bmod m(x) \big) < N-t.
\end{equation}
Then $\Lambda(x)$ is a multiple of $\Lambda_f(x)$. Conversely,
$\Lambda(x)=\Lambda_f(x)$ is a polynomial of the smallest degree
that satisfies (\ref{eqn:OurPRCKey}).
\end{theorem}

It follows that the decoding procedure of
Section~\ref{sec:Decoding} works also for polynomial remainder
codes, except that $n$, $k$, and $\Lambda_e(x)$ are replaced by
$N$, $K$, and $\Lambda_f(x)$, respectively. Moreover, $C(x)$ can
still be recovered from $\Lambda(x)$ by means of
(\ref{eqn:RecoverCxByMultDiv}) \cite{YuLoeliger:prc:ArXiv2012}.

\newcommand{\COM}{IEEE Trans.\ Communications}
\newcommand{\COMMag}{IEEE Communications Mag.}
\newcommand{\IT}{IEEE Trans.\ Information Theory}
\newcommand{\JSAC}{IEEE J.\ Select.\ Areas in Communications}
\newcommand{\SP}{IEEE Trans.\ Signal Proc.}
\newcommand{\SPMag}{IEEE Signal Proc.\ Mag.}
\newcommand{\ProcIEEE}{Proceedings of the IEEE}


\begin{thebibliography}{99}

\bibitem{Reed}
I.~S.~Reed and G.\ Solomon, ``Polynominal codes over certain
finite fields,'' \emph{ J.~SIAM}, vol.~8, pp.\ 300--304, Oct.\
1962.

\bibitem{Berlekamp}
E.~R.~Berlekamp, ``Algebraic Coding Theory.'' New York:
McGraw-Hill, 1968.

\bibitem{Sugiyama}
Y.~Sugiyama, M.~Kasahara, S.~Hirasawa, and T.~Namekawa, ``A method
for solving key equation for decoding Goppa codes,''
\emph{Information and Control}, vol.\ 27, pp.\ 87--99, 1975.

\bibitem{Blahut}
R.~E.~Blahut, ``Algebraic Codes for Data Transmission.'' Cambridge
University Press, Cambridge, UK, 2003.

\bibitem{Roth}
R.~M.~Roth, \emph{Introduction to Coding Theory.}
New York: Cambridge University Press, 2006.

\bibitem{Massey}
J.~L.~Massey, ``Shift-register synthesis and BCH decoding,''
\emph{ \IT}, vol.~15, pp.\ 122-127, May\ 1969.

\bibitem{Shiozaki}
A.~Shiozaki, ``Decoding of redundant residue polynomial codes
using Euclid's algorithm,'' \emph{ \IT}, vol.\ 34, pp.\
1351--1354, Sep.\ 1988.

\bibitem{Gao}
S. Gao, ``A new algorithm for decoding Reed-Solomon codes,'' in
\emph{Communications, Information and Network Security}, V.
Bhargava, H. V. Poor, V. Tarokh, and S.Yoon, Eds. Norwell, MA:
Kluwer, 2003, vol. 712, pp. 55-68.


\bibitem{Stone}
J.~J.~Stone, ``Multiple-burst error correction with the Chinese
Remainder Theorem,'' \emph{ J.~SIAM}, vol.~11, pp.\ 74--81, Mar.\
1963.


\bibitem{YuLoeliger}
J.-H.~Yu and H.-A.~Loeliger, ``On irreducible polynomial remainder
codes,'' \emph{IEEE Int. Symp. on Information Theory, Saint
Petersburg, Russia, July 31--Aug. 5, 2011}.

\bibitem{YuLoeliger:prc:ArXiv2012}
J.-H.~Yu and H.-A.~Loeliger, ``On polynomial remainder codes,''
http://arxiv.org/abs/1201.1812.


\bibitem{Dornstetter}
J.~L.~Dornstetter, ``On the equivalence between Berlekamps's and
Euclids's algorithms,'' \emph{ \IT}, vol.~33, pp.\ 428-431, May\
1987.

\bibitem{Fitzpatrick}
P.~Fitzpatrick, ``On the key equation,'' \emph{ \IT}, vol.~41,
pp.\ 1290-1302, Sep.\ 1995.

\bibitem{Heydtmann}
A.~E.~Heydtmann and J.~M.~Jensen, ``On the equivalence of the
Berlekamp-Massey and Euclidean algorithms for decoding,''
\emph{\IT}, vol.~46, pp.\ 2614-2624, Nov.\ 2000.


\end{thebibliography}
\end{document}